# Conformational Transition of H-shaped Branched Polymers

Ashok Kumar Dasmahapatra* and Venkata Mahanth Sanka

Department of Chemical Engineering, Indian Institute of Technology Guwahati, Guwahati – 781039, Assam, India

* Corresponding author: Phone: +91-361-258-2273; Fax: +91-361-258-2291; Email address: akdm@iitg.ernet.in

## ABSTRACT

We report dynamic Monte Carlo simulation on conformational transition of H-shaped branched polymers by varying main chain (backbone) and side chain (branch) length. H-shaped polymers in comparison with equivalent linear polymers exhibit a depression of theta temperature accompanying with smaller chain dimensions. We observed that the effect of branches on backbone dimension is more pronounced than the reverse, and is attributed to the conformational heterogeneity prevails within the molecule. With increase in branch length, backbone is slightly stretched out in coil and globule state. However, in the pre-collapsed (cf. crumpled globule) state, backbone size decreases with the increase of branch length. We attribute this non-monotonic behavior as the interplay between excluded volume interaction and intra-chain bead-bead attractive interaction during collapse transition. Structural analysis reveals that the inherent conformational heterogeneity promotes the formation of a collapsed structure with segregated backbone and branch units (resembles to “sandwich” or “Janus” morphology) rather an evenly distributed structure comprising of all the units. The shape of the collapsed globule becomes more spherical with increasing either backbone or branch length.

## I. INTRODUCTION

Branched polymers are commodious in preparing tailor-made materials for various applications such as in the area of catalysis[1], nanomaterials[2], and biomedicines[3]. Properties of branched polymers can be tuned by controlling number, length and distribution of branches along the backbone chain. For example, to achieve low crystalline polyethylene, short branches of α-olefin, largely butene, hexene or octene, are incorporated along the backbone chain[4]. In bulk, short branches do not crystallize; rather, they act as defects in the polymer crystal resulting less crystalline polymer, which is useful for specific applications. Understanding of branching characteristics on phase behavior and chain dynamics would enable to synthesize polymers with precise branching pattern (viz., number of branches and their distribution along the backbone chain).

H-shape branch polymer is a special type of branch polymer, which has one backbone with two branches originated from each of the ends of the backbone (see Figure 1 for a schematic representation). H-shape polymer may be visualized as a miniature version of a long branched polymer like LDPE or LLDPE. Advancement of synthetic chemistry paves the pathway to synthesize polymers with complex architectures, such as star, dendrimer[5-8] and branched polymers with precisely controlled branching[9]. Anionic polymerization has been used to prepare various H-shape branch polymers[10-15]. Controlled chlososilane coupling chemistry has been coupled with the anionic polymerization technique to tune the structure of H-polymer with almost monodisperse molecular weight distribution of poly(butadiene)[10,12] with a wide variety of branch lengths. Reversible addition-fragmentation chain transfer (RAFT) coupled with cation ring opening polymerization (CROP) has been applied to synthesize H-shape copolymer of polystyrene as backbone and poly(1,3-dioxepane) as branches[16]. Ring opening metathesis polymerization has been used to synthesis H-shaped copolymer with telechelic poly(cis-cyclooctene) as backbone and D,L-lactide as branch[17].

H-shape polymers, due to its branching, would behave differently than the linear polymers. Branches have a free dangling end, which contributes more conformational entropy towards free energy than the backbone. Therefore, an energetically heterogeneity exists within the molecule, which influence the conformational properties of H-polymers. In dilute solution, polymers exist in a swollen coil state under good solvent condition. On deteriorating solvent quality (viz., lowering temperature) conformational change occurs from an expanded coil state to the compact globule state via theta. In good solvent conditions, H-shaped polymers show a smaller chain dimension compared to linear polymers with comparable degree of polymerization[18,19]; although they follow the same scaling relation as linear polymers with a scaling exponent ~ 0.6[13]. In the theta state, according to Flory[20], polymer chains behave ideally and mean square radius of gyration scales with *N*, where *N* is the degree of polymerization. The chain dimension and theta temperatures are dictated by the topological features of the polymers. Roovers and Toporowski[11] have demonstrated that the behavior of H-shaped polymer lie between three- and four-arm star polymers. They have observed that the second virial coefficient is positive and theta temperature of H-shaped polystyrene in cyclohexane is lower than that of linear polystyrene under identical condition. Monte Carlo simulation suggests that the value of second and third virial coefficient of H-polymers are higher than the linear polymers with equivalent molecular weight[21]. It has also been observed that H-shaped polymer exhibits a bulk rheological pattern, which is distinctly different than linear polymer[22-25].

Although, the bulk behavior of H-polymers is fairly understood, solution behavior of H-polymers is not well explored. In this paper, we describe simulation results on collapse transition of H-shaped polymers with varying length of backbone and branches. We observe that the dimensions of H-polymers are smaller than the corresponding linear polymers. We have analyzed the effect of branching on the collapse behavior by varying branch length (*N2*) while keeping backbone length (*N1*) constant, and by varying *N1* keeping *N2* constant. Variation in *N2* would represent short arm to long arm branches, whereas variation in *N1* represent the degree of branching along the backbone chain in a branch polymer. We observe that the effect of branch length on backbone conformation is more pronounced than the reverse. Structurally, the collapsed globule resembles to “sandwich” or “Janus” morphology, wherein backbone and branch units are well segregated.

We organize our paper as follows. In the section II, we describe our model and simulation technique. We discuss our key results in the section III followed by summary in the section IV.

## II. MODEL AND SIMULATION TECHNIQUE

We use dynamic Monte Carlo (DMC) simulation on a simple cubic lattice. We model an H-shaped branched molecule as a molecule containing one backbone with two number of branches (or arms) originating from either end of the backbone. For an H-polymer with each branch containing *N2* number of repeat units, and backbone containing *N1* number repeat units, the total number of unit becomes $N = N1 + 4N2$ (Figure 1). We use *N1_N2_N* as the nomenclature for H-polymers with various values of backbone (*N1*) and branch length (*N2*).

A polymer chain is represented by connecting successive lattice sites on a cubic lattice of size 200 × 200 × 200. Occupied lattice sites are monomer (*m*) and vacant sites represent solvent molecules (*s*). We employed single site bond fluctuation method[26] along with periodic boundary condition to simulate the chain molecule in the lattice. To give further details, we begin our simulation by selecting a monomer randomly and attempt to move to the nearest lattice site. A strict self avoiding walk model chain has been used to implement excluded volume interaction. One lattice site is occupied by only one unit and no bond crosses with any other bonds. We have employed the Metropolis sampling[27] to sample new conformation with a probability $p = \exp\left(-\Delta E/kT\right)$, where $\Delta E$ is the change in energy in a given MC move, *k* is the Boltzmann constant and *T* is the temperature in K. The new conformation is accepted if $p \geq r$, where *r* is a random number, $0 \leq r \leq 1$, generated by using random number generator, MT19937[28]. Interaction between monomer and solvent has been modeled by quasi-chemical approximation, incorporating exchange energy between monomer and solvent. Thus, the change in energy in an MC move is modeled as: $\Delta E = \Delta N_{ms} B_{ms}$, where, $\Delta N_{ms}$ are the net change in the number of contacts between *m* and *s* units; $B_{ms}$ are the exchange energies (normalized by *kT*) between *m*-*s* contacts respectively. We take $B_{ms} = B$, which is equivalent to the Flory's $\chi$ parameter and inversely proportional to temperature (viz. $B \sim T^{-1}$). At each value of *B*, we equilibrate the sample for large number of Monte Carlo steps (MCS), and calculate thermodynamic and structural properties over an equal number of MCS. *N* number of attempted MC move is defined as one MCS. We equilibrate the sample system at $B = 0$ ($T = \infty$, athermal state) and progressively cooled the system by increasing *B* in steps of 0.002. We have not observed any hysteresis during cooling (increasing *B*) and heating (decreasing *B*) runs. Hence, our simulations represent equilibrium slow cooling experiments. To monitor the transition from coil to globule, we calculate radius of gyration as: $R_g^2 = \sum_{i=1}^{K} \left(r_i - r_{cm}\right)^2 \Big/ K$, where, $r_i$ is coordinates of *i*-th unit and $r_{cm}$ is center of mass defined as: $r_{cm} = \sum_{i=1}^{K} r_i \Big/ K$. *K* represents number of units taken into consideration for the calculations of $R_g^2$: $K = N1$, $N2$ and $N$ to calculate $\langle R_g^2 \rangle$ for the backbone, branch and entire molecule respectively.

We also calculate mean square end to end distance, $\langle R^2 \rangle$ of H-polymers. We express, $\langle R^2 \rangle$ as the branch-to-branch distance for backbone and branch-to-end distance for branches. For branch, the value of $\langle R^2 \rangle$ and $\langle R_g^2 \rangle$ has been estimated as an average over four branches.

We have also estimated the evolution of shape of the polymer chain during collapse transition by calculating principal components of the shape tensor. The shape tensor $S^2$ is defined as[29,30]:

$$S^2 = \begin{bmatrix} S_{xx} & S_{xy} & S_{xz} \\ S_{yx} & S_{yy} & S_{yz} \\ S_{zx} & S_{zy} & S_{zz} \end{bmatrix}$$

where, $S_{xx} = \frac{1}{N}\sum_{i=1}^{N}\left(x_i - x_{cm}\right)^2$ and $S_{xy} = \frac{1}{N}\sum_{i=1}^{N}\left(x_i - x_{cm}\right)\left(y_i - y_{cm}\right)$, and $x_i$ and $x_{cm}$ are the *x*-coordinate of *i*th monomer and center of mass, respectively.

The eigenvalues of the shape tensor yield the three principal moments of $S^2$ and represent the shape of the molecule. The eigenvalues are represented as $S_i^2$. To characterize the shape of the molecule, we define the asphericity parameter, $\delta^*$ by the following formula[30]:

$$\delta^* = \frac{\left(S_1^2 - S_2^2\right)^2 + \left(S_2^2 - S_3^2\right)^2 + \left(S_3^2 - S_1^2\right)^2}{2\left(S_1^2 + S_2^2 + S_3^2\right)^2} \qquad (1)$$

For an isotropic structure, $\delta^*$ equals to zero.

## III. RESULTS AND DISCUSSION

We begin to describe our simulation results by comparing a linear and H-shaped branched polymer with $N$ = 512. Subsequently, we present results for H-shaped branched polymers with varying length of backbone ($N1$) and branch ($N2$), for a series of combinations (constant and variable $N$), to elucidate the effect of branching on the collapse behavior of H-polymers.

### III.A. Comparison between Linear and H-shaped Polymer

We simulate a linear polymer chain with $N$ = 512, and two H-shaped branched polymers, with different combination of backbone and branch length: 256_64_512 and 128_96_512. Figure 2 represents the variation of $\left\langle R_g^2 \right\rangle$ as a function of $B$ for all three polymers. At $B = 0$ ($T = \infty$, athermal state), molecules posses an expanded coil conformation and H-polymers have lower dimension in comparison with the linear polymer. The lower dimension of branch polymer in comparison with the linear polymer is characterized by branching parameter, given by, $g = \left\langle R_g^2 \right\rangle_{br} \Big/ \left\langle R_g^2 \right\rangle_l$ . We calculate the branching parameter from our simulation, and observe that the values are in close agreement to that of calculated from the following theoretical equation[15,18]:

$$g_{th} = \lambda^3 + 3\lambda^2(1-\lambda) + \frac{9}{4}\lambda(1-\lambda)^2 + \frac{5}{8}(1-\lambda)^3 \qquad (2)$$

where, $\lambda$ is the number fraction of backbone units (in our model, $\lambda = N1/N$). For 256_64_512 and 128_96_512 H-polymers, $\lambda$ = 0.5 and 0.25 respectively, and the corresponding $g_{th}$ are 0.85 and 0.74 respectively. Simulation results show that the value of $g$ is 0.86 and 0.72, which are in close agreement to the theoretical values. It is to be noted that equation 2 is valid for H-polymers in melt or under theta condition (in case of solution). However, in the good solvent condition, our simulation results match well with the theoretical value. Simulation of a series of H-polymers with various combination of $N1$ and $N2$, results in the same range of $g$, comparable to the theoretical prediction. Table 1 summarizes simulated systems; estimated values of $g$ form our simulation, and calculated values of $g$ ($g_{th}$) using equation 2 for various values of $\lambda$. We now check the values of $g$ with $g_{th}$ under theta condition.

According to Flory[20], a polymer chain behaves ideally at theta state, following the scaling relation of $\langle R_g^2 \rangle \approx N$. Thus, the ratio of $\langle R_g^2 \rangle / N$ would be a constant at theta state. We employ this criteria to estimate the theta point[29]: we simulate polymer chains of different length (for linear chain, $N$ = 64, 128, 256 and 512; and for H-polymer, 32_8_64, 64_16_128, 128_16_256 and 256_64_512) and estimated the theta value as the value of $B$ at which $\langle R_g^2 \rangle / N$ is almost constant for all the chain length simulated. Strictly speaking, a true theta is only realizable at $N \to \infty$. For a finite chain length system, we find rather a "theta region" that approaches the true unique theta point[31] as $N \to \infty$. For linear chain (Figure 3a), we estimate the theta region as: $B$ = 0.027 ± 0.001 and $\langle R_g^2 \rangle / N$ = 0.4 ± 0.01. For H-polymer (Figure 3b), the corresponding values are: $B$ = 0.029 ± 0.001 and $\langle R_g^2 \rangle / N$ = 0.34 ± 0.01. The results clearly exhibit that the chain size at the theta state is lower for H-polymer compared to that of linear polymer. The theta value of $B$ of H-polymers is higher than linear polymer, which signifies a depression of theta temperature ($B \sim 1/T$) for H-polymer, which is in accord with the literature[11]. We calculate $g$ at the theta point and observe a close agreement with the theoretical value: $g = \langle R_g^2 \rangle_{br} / \langle R_g^2 \rangle_l$ = 0.34/0.4 = 0.85 for $\lambda$ = 0.5 (see Table 1).

With increasing value of $B$ (beyond theta), the effective *m-m* attractive interaction energy increases and it counterbalances the *m-s* repulsive interaction. As a result, the chain starts shrinking and finally reaches to a compact globule state at $B \geq 0.6$. H-shaped polymers too follow the similar trend to that of linear polymer. Throughout the temperature range (beyond theta), we observe that H-polymers possess equal dimension to that of linear polymers (Figure 2 and Figure 3a and 3b). The presence of two branch points in the H-shaped polymer does not influence the overall size and shape of the collapsed globule. Beyond theta, the free energy is dominated by the enthalpic contribution rather than entropic, and therefore, the presence of branch points does not make any difference in the size of the globule. However, the value of $\langle R_g^2 \rangle / N$ decreases with increasing $N$ (higher value for smaller chains), which may be attributed to the improper packing for low molecular weight polymer; the compactness of the structure is achieved in high molecular weight polymer. To elucidate further, we define a density of the chain as number of units, $N$ divided by $\langle R_g^2 \rangle^{3/2}$, a measure of the chain volume ( $N/\langle R_g^2 \rangle^{3/2}$ ) and estimated as a function of $B$ (Figure 4). The value of $N/\langle R_g^2 \rangle^{3/2}$ increases with increasing $B$ from coil state to reach almost a saturated value in the globule state for all the chain length simulated. The magnitude of density increases with increase in $N$ and in the globule state, higher value of $N$ results in the formation of globules with higher density (Figure 4). A dense globule is resulted from the enhanced *m-m* interaction, which overcomes excluded volume interaction. As presented in Figure 2, the effect of branching is more pronounced in the coil state than in the globule state for a constant $N$ system (linear and H-polymers). In the following, we discuss the effect of branching on collapse behavior of H-polymers, by considering various combinations of *N1* and *N2* (viz., degree of branching and branch length respectively).

### III.B. Effect of branch length on backbone conformation (Constant *N1*, variable *N2*)

To investigate the effect of branch length on backbone conformation, we vary branch length (*N2*) for a fixed backbone length (*N1*). We estimate $\langle R^2 \rangle$ and $\langle R_g^2 \rangle$ for backbone and branches for every combination of *N1* and *N2*. Figure 5 represents the variation of these values with $B$, from coil to globule state for $N1 = 64$, with $N2 = 4$, 8, 16, 32, 64, 80 and 96. In the coil state, with the increase in value of *N2*, the value of $\langle R^2 \rangle_{N1}$ (Figure 5a) and $\langle R_g^2 \rangle_{N1}$ (Figure 5b) increases monotonically, which is counter-intuitive: for a constant *N1*, value of $\langle R^2 \rangle_{N1}$ and $\langle R_g^2 \rangle_{N1}$ would be expected to be same. As the branch length increases, backbone is slightly stretched out due to the attainment of coil conformation (cf. Gaussian chain) of branches. Branches (with one branch point) are entropically in a favorable condition than backbone (with two branch points). As a result, branches are more solvated than the backbone. We have analyzed branch conformational behavior for a fixed value of *N1*. We found that the branches follow a self-avoiding walk (SAW) scaling ($\langle R_g^2 \rangle_{N2} \sim N2^{2v}$) with a scaling exponent, $v \sim 0.6$. These results are in line with the experimental and simulation results for highly branched polymers like polymer brush[32-35], wherein side chains increases the stiffness of the backbone.

As the solution is cooled further (below theta, $B > 0.03$), molecules shrink to a pre-collapsed state, where monomers are still solvated in the solvent. Few aggregates of monomers are observed, which eventually coalesce to form a compact globule on further decreasing temperature. This state has been described as the molten globule state[36]. We observe a decreasing value of $\langle R^2 \rangle_{N1}$ and $\langle R_g^2 \rangle_{N1}$ with increasing value of *N2*. Temperature below theta, the excluded volume effect is being counterbalanced by the enhanced *m-m* attractive interaction. Due to decrease in excluded volume (with enhanced *m-m* attractive interaction), the average volume per monomer decreases with increasing *N* (see Figure 3 for example). The monomers of *N1* chain are in contact with more number of neighboring monomers (from both *N1* and *N2*). This enhanced monomer contacts facilitates to possess a reduced volume of *N1* chain, and is reflected in the lower value of $\langle R^2 \rangle_{N1}$ and $\langle R_g^2 \rangle_{N1}$. A free dangling end facilitates branches to be more solvated than backbone. As a result, backbone length follows a decreasing trend with increasing *N2* (also increasing *N*). $\langle R_g^2 \rangle$ of the entire chain (normalized by *N*, $\langle R_g^2 \rangle / N$) also follow a similar trend (Figure 5c) with increasing value of *N*. We observe a similar results for other cases where we simulate H-polymers with *N1* = 32 (*N2* = 4, 8, 16, 32, 48 and 64), 128 (*N2* = 4, 8, 16, 32, 64 and 96) and 256 (*N2* = 4, 8, 16, 32 and 64). Figure 6 summarizes the variation of $\langle R^2 \rangle_{N1}$ and $\langle R_g^2 \rangle_{N1}$ with *N2* for *N1* = 32, 64, 128 and 256 at crumpled globule state ($B = 0.036$).

On further cooling ($B \geq 0.05$), *m-m* attractive interaction dominates over excluded volume interaction. As a result, the chain molecule collapses into a compact globule state. In the collapsed state, we observe that with increase in *N2*, $\langle R^2 \rangle_{N1}$ and $\langle R_g^2 \rangle_{N1}$ again increase monotonically. For H-polymers, with short branch length, polymer collapse is mainly driven by the collapse of backbone without getting influenced by branches. As the branch length increases, collapse of branch may happen independently of backbone (branches follow a SAW scaling, behaves like Gaussian chains). On the other hand, four branches may collapse co-operatively at higher value of *N2*. Further, monomer density near the branch point is more than the rest of the chain. Collapse of branches is facilitated by the enhanced monomer density at the branch point. As the branch points are crowded by the collapse of branch units, the collapse of backbone to attain a compact state becomes difficult. As a result, backbone dimension increases slightly with increase in branch length (*N2*). We observe a similar trend for other cases where we simulate H-polymers with *N1* = 32 (*N2* = 4, 8, 16, 32, 48 and 64), 128 (*N2* = 4, 8, 16, 32, 64 and 96) and 256 (*N2* = 4, 8, 16, 32 and 64). Figure 7 summarizes the variation of $\langle R^2 \rangle_{N1}$ and $\langle R_g^2 \rangle_{N1}$ with *N2* for *N1* = 32, 64, 128 and 256 at the collapsed state ($B = 0.1$).

### III.C. Varying backbone and branch length – keeping total units constant

We simulate H-polymers where we increase *N1* and decrease *N2* such that *N* remains constant. In Figure 2 we have presented results for two H-polymers with *N* = 512, with different combination of *N1* and *N2*. We have seen that $\langle R_g^2 \rangle$ of 256_64_512 is higher than that of 128_96_512, in the coil state (Figure 2). However, as the solution is cooled further, ($B \geq 0.04$), the value of $\langle R_g^2 \rangle$ appears to be same and it continues till the collapsed state. It appears that even though the total number of units (*N*) remains constant in the molecule, the change in the backbone (*N1*) and branch (*N2*) length dramatically change the conformational behavior of the entire chain, especially in the coil state. Figure 8 shows the results for three H-polymers with *N* = 160 and 192. For the case with varying *N1* and *N2* (*N* constant), the difference in chain dimension ($\langle R_g^2 \rangle$) exists in the coil state, but in the globule state, the difference diminishes to almost zero. Once the temperature crosses theta, the coil size of all the chain behaves almost identically. The entropic contribution towards free energy decreases due to the decrease in branch length (*N2*); therefore, the difference in the value of $\langle R^2 \rangle$ and $\langle R_g^2 \rangle$ in the coil state is not too large. As we can see from Figure 2 and Figure 8, $\langle R_g^2 \rangle$ is higher for H-polymer with shorter branch length than that of longer branch length. For shorter branch length, molecule appears more like a linear chain; whereas, as the branch length increases (and backbone length decreases), $\langle R_g^2 \rangle$ decreases.

### III.D. Backbone and Branch are of equal length

We simulate H-polymers with equal value of *N1* and *N2*. Figure 9a,b presents our simulation data for *N1* = 32 and 64 respectively. We observe that throughout the range of *B* (viz., from coil to globule state), branches possess higher dimension than backbone. The presence of a free dangling end in branches facilitates to explore larger conformational space than backbone. As a result, branches are more solvated than backbone. As the chain collapses, backbone units collapse before branch units owing to the enhanced *m-m* interaction. Therefore, $\langle R_g^2 \rangle$ of branches is higher than that of backbone.

### III.E. Effect of backbone length on branch conformation

We keep branch length constant and vary backbone length to investigate the effect of backbone on conformational behavior of branches. Variation of backbone length (*N1*) may be correlated to the degree of branching (average backbone length between two nearest branch points) in highly branched polymers like LDPE or LLDPE. Figure 10a and b present the simulation results for *N2* = 32 (*N1* = 32, 64, 128 and 256) and 64 (*N1* = 64, 128 and 256) respectively. In coil state, with increasing the length of the backbone (*N1*) for a given branch length (*N2*), the value $\langle R_g^2 \rangle_{N2}$ remains almost constant. In the coil state, excluded volume interaction (viz., entropic contribution) dominates over *m-m* attractive interaction (viz., enthalpic contribution), and hence, there is hardly any effect of the backbone length on branch conformation. We have analyzed backbone conformational behavior for a fixed value of *N2*. We observe that the backbone too follows a SAW scaling ($\langle R_g^2 \rangle_{N1} \sim N1^{2\nu}$) with a scaling exponent, $\nu \sim 0.57$. The scaling behavior clearly shows that the influence of branches on backbone is more pronounced than the reverse. In contrast to the results presented in Figure 5 (size of the backbone increases with increase of branch length), we observe that size of the branches is almost independent of backbone length.

As the solution is cooled further below theta point ($B > 0.03$), we observed an increase in the coil size of branches with increasing *N1*. As temperature decreases (viz., *B* increases), enthalpic interaction dominates over entropic interaction. With increase in backbone length, distance between branches increases and the probability of co-operative collapse between branches decreases. Branches with a free dangling end facilitate in possessing higher conformational entropy compared to backbone. Therefore, in decreasing temperature, collapse of backbone precedes the collapse of branches. As a result, backbone occupies the inner part of the globule, surrounded by branches, and results a larger dimension of branches with increasing backbone length. Below, we present detail structural analysis including the shape of the molecule from our simulation.

### III.F. Structural Analysis

The nature of collapse transition presented in Figure 2, 3 and 5 reveals that the collapse happens via a two-stage process: collapse of random coil to a crumpled globule state ($B \sim 0.036$), and then from crumpled globule to a collapsed globule state at $B \geq 0.6$. Collapse is initiated by the formation of a dense core largely dominated by backbone units followed by the collapse (or reeling in) of branch units. We have seen that in the pre-collapsed (viz., crumpled globule) state ($B \sim 0.036$), the dimension of the backbone decreases with increasing branch length (Figure 5 and Figure 6). In most cases, branches possess an expanded conformation than backbone, signifying a smaller dimension of backbone in comparison with branches. Figure 11 represents the snapshots from simulation at $B = 0.036$ for 64_16_128, 64_32_192, 64_48_256, 64_64_320 and 64_80_384. It is to be noted that we simulate H-shaped homopolymer (backbone and branch units are chemically identical). We have used black and magenta symbol to represent backbone and branch units respectively, for better visualization. As the solution is cooled further, the crumpled globule collapses to a compact globule structure. In the collapsed state ($B = 0.1$), backbone dimension slightly increases with increasing *N2* for a given value of *N1* (Figure 5 and Figure 7). When we measure $\left\langle R_g^2 \right\rangle_{N2}$, we found that branches possess larger dimension compared to that of backbone (Figure 9), except for very short branches. The presence of branches inhibits the formation of a globule where branch and backbone units may evenly be distributed. As a result, branches remain at the surface of the globule and possess a larger dimension than backbone. Different morphological pattern of collapsed globules have been observed depending on the value of *N1* and *N2*. In some cases, the final structure resembles to "sandwich" (Figure 12: 64_4_80, 64_16_128, 128_32_256, and 128_96_512), where backbone is flanked by branches from either end, or "Janus" (Figure 13: 64_32_192, 64_64_320, 64_80_384, 64_96_448) morphology, where backbone monomers are almost separated from branch monomers. In these types of structures, backbone and branch units are almost segregated. These type of segregated globule structures are usually observed in the collapse transition of heteropolymers with "sticky" comonomers[29,37], where co-monomers form the core surrounded by monomers leading to a core-shell morphology. The segregated structure (vis., core-shell) is primarily driven by the energetic heterogeneity within the molecule (viz., monomers and comonomers are chemically different and hence interactions are different). However, in the present case (H-shape homopolymer), we see the formation of equivalent segregated globule structures due to the presence of conformational heterogeneity (originated from the difference in entropy of backbone and branch) within the molecule.

***Shape analysis:*** We calculate asphericity factor, $\delta^*$ (equation 1), to monitor change in shape along with the structural transformation of H-polymers during collapse transition. Figure 14a and Figure 14b illustrate the variation of $\delta^*$ with increasing $B$ for a series of linear and H-polymers respectively. Shape of the molecule becomes more spherical as it is cooled from coil to globule state, where it assumes a compact geometry. In the coil state, $\delta^*$ remains almost same for all $N$ (Figure 14a and b). In the globule state, $\delta^*$ decreases with increasing $N$.

For a given value of backbone ($N1$ = 64), with increasing branch length ($N2$ = 4, 8, 16, 32, 64, 80 and 96), $\delta^*$ decreases throughout the range of $B$ (Figure 15a). For short branch length, as we have seen (section III.B) that the collapse is driven primarily by the backbone, and the collapse happens in a manner similar to a linear polymer. As $N2$ increases ($N$ also increases), the entire H-polymer approximates to a spherical shape, and the value of $\delta^*$ decreases. We observe a similar behavior for other values of $N1$ with a series of $N2$ value. Figure 15a (inset) presents the value of $\delta^*$ with $N2$ for a series of $N1$ at the collapsed state ($B$ = 0.1). From this figure, it is clear that with the increase in $N2$ (increase in $N$), the shape of the globule becomes more spherical.

For a given value of $N2$, with increasing value of $N1$, different scenario is observed in coil and globule state: in the coil state, $\delta^*$ increases with increasing $N1$; and in the globule state, $\delta^*$ decreases with increasing $N1$ (Figure 15b). In the coil state, with increasing $N1$, the molecule approximates to a linear chain and as results, shape deviates from being spherical (looks like a dumbbell). In the globule state, with increase in $N$ (by increasing either $N1$ or $N2$) shape of the globule becomes more compact and spherical, which has also been observed for linear chains. For a constant $N$ system (by varying $N1$ and $N2$), the shape of the collapsed globule remains similar to that of linear polymer. However, in the coil state, shape of H-polymers differs from the linear chain (Figure 15c). In the coil state, 128_96_512 H-polymer exhibits lower value of $\delta^*$ than the others. It appears that the closer the value of $N1$ and $N2$, more spherical shape it may possess. Longer chains produce more compact globule structure, which assumes a better spherical shape, than shorter ones. Although branch points influences the collapse transition, it appears from the above analysis that the globule size and shape are not sensitive to the "branchingness" of H-polymers, and behaves similar to a linear polymer for a system with equal *N*.

## IV. CONCLUSIONS

Branching characteristics plays a crucial role in determining solution behavior of branched polymers. The number and length of the branches dictate the overall conformations of branched polymers. In this work, we have presented simulation results for a series of H-shaped branched polymers, by varying the length of backbone and branches, and found that the relative ratio of the length of these two plays a crucial role in deciding the conformational behavior in solution. H-polymers, in comparison with equivalent linear polymer, have smaller chain dimension from coil to globule state. We have also observed a depression in theta temperature for H-polymer in comparison with linear polymers. Presence of branch points influences the conformational behavior of both backbone and branches. For a given value of *N1*, $\langle R_g^2 \rangle_{N1}$ increases with increase of *N2* in the coil and globule state, but follows a reverse trend in the crumpled globule state. We have interpreted as the interplay between excluded volume repulsive interaction and *m-m* attractive interaction. For a given value of *N2*, in the coil state, there is no effect on the branch size in increasing value of *N1*, but in the globule state chain dimension increases with increasing *N1*. Scaling analysis shows that the branches (for a constant *N1*) follows a scaling relation with a scaling exponent, $v \sim 0.603$, whereas backbones for a constant branch follow a scaling relation with a scaling exponent, $v \sim 0.57$. This variation in scaling exponent suggests that the effect of branching is more pronounced for backbone conformation than branches. Structural analysis reveals that the globules possess a segregated distribution rather than an even distribution of backbone and branch units. Conformational heterogeneity leads to the formation of such morphology, and depending on the length of backbone and branches, globule structure vary from "sandwich" (Figure 12) to "Janus" (Figure 13) morphology. Analysis on shape factor reveals that with increasing either the length of backbone or branches, the shape of the collapsed globule appears to be more spherical. Our results on the conformational behavior of H-polymers in dilute solution would enable in gaining valuable insight to understand the internal segmental dynamics, which may have a profound effect on rheological behavior of branched polymers.

REFERENCES

[1] D. A. Tomalia and P. R. Dvornic, Nature **372**, 617 (1994).

[2] K. Landskron and A. O. Geoffrey, Science **306**, 1529 (2004).

[3] R. Esfand and D. A. Tomalia, DDT **6**, 427 (2001).

[4] C. Baig, O. Alexiadis, and V. G. Mavrantzas, Macromolecules **43**, 986 (2010).

[5] K. Adachi, H. Irie, T. Sato, A. Uchibori, M. Shiozawa, and Y. Tezuka, Macromolecules **38**, 10210 (2005).

[6] H. R. Kricheldorf and H. H. Thießen, Polymer **46**, 12103 (2005).

[7] H. R. Kricheldorf and T. Stukenbrock, Polymer **38**, 3373 (1997).

[8] M. A. Tasdelen, M. U. Kahveci, and Y. Yagci, Prog. Polym. Sci. **36**, 455 (2011).

[9] J. C. Sworen, J. A. Smith, J. M. Berg, and K. B. Wagener, J. Am. Chem. Soc. **126**, 11238 (2004).

[10] M. S. Rahman, H. Lee, X. Chen, T. Chang, R. Larson, and J. Mays, ACS Macro Lett. **1**, 537 (2012).

[11] J. Roovers and P. M. Toporowski, Macromolecules **14**, 1174 (1981).

[12] S. Perny, J. Allgaier, D. Cho, W. Lee, and T. Chang, Macromolecules **34**, 5408 (2001).

[13] A. Hakiki, R. N. Young, and T. C. B. McLeish, Macromolecules **29**, 3639 (1996).

[14] X. Chen, M. S. Rahman, H. Lee, J. Mays, T. Chang, and R. Larson, Macromolecules **44**, 7799 (2011).

[15] M. S. Rahman, R. Aggarwal, R. G. Larson, J. M. Dealy, and J. Mays, Macromolecules **41**, 8225 (2008).

[16] J. Liu and C.-Y. Pan, Polymer **46**, 11133 (2005).

[17] L. M. Pitet, B. M. Chamberlain, A. W. Hauser, and M. A. Hillmyer, Macromolecules **43**, 8018 (2010).

[18] D. S. Pearson and V. R. Raju, Macromolecules **15**, 294 (1982).

[19] J. Ramos, L. D. Peristeras, and D. N. Theodorou, Macromolecules **40**, 9640 (2007).

[20] P. J. Flory, *Principles of Polymer Chemistry* (Cornell University Press, Ithaca, New York, 1953).

[21] K. Shida, A. Kasuya, K. Ohno, Y. Kawazoe, and Y. Nakamura, J. Chem. Phys. **126**, 154901-1 (2007).

[22] T. C. B. McLeish, J. Allgaier, D. K. Bick, G. Bishko, P. Biswas, R. Blackwell, B. Blottière, N. Clarke, B. Gibbs, D. J. Groves, A. Hakiki, R. K. Heenan, J. M. Johnson, R. Kant, D. J. Read, and R. N. Young, Macromolecules **32**, 6734 (1999).

[23] T. C. B. McLeish, Macromolecules **21**, 1062 (1998).

[24] J. Roovers, Macromolecules **17**, 1196 (1984).

[25] A. Santamaria, Materials Chemistry and Physics **12**, 1 (1985).

[26] A. K. Dasmahapatra and G. D. Reddy, Polymer **54**, 2392 (2013).

[27] N. Metropolis, A. W. Rosenbluth, M. N. Rosenbluth, A. H. Teller, and E. Teller, J. Chem. Phys. **21**, 1087 (1953).

[28] T. Nishimura and M. Matsumoto, http://www. math. sci. hiroshima-u. ac. jp/~m-mat/MT/emt. html (2002).

[29] A. K. Dasmahapatra, G. Kumaraswamy, and H. Nanavati, Macromolecules **39**, 9621 (2006).

[30] A. Sikorski and P. Romiszowski, J. Chem. Phys. **109**, 6169 (1998).

[31] C. Domb, Polymer **15**, 259 (1974).

[32] A. Yethiraj, J. Chem. Phys. **125**, 204901-1 (2006).

[33] K. Fischer and M. Schmidt, Macromol. Rapid Commun. **22**, 787 (2001).

[34] M. Gerle, K. Fischer, S. Roos, A. H. E. Müller, M. Schmidt, S. S. Sheiko, S. Prokhorova, and M. Möller, Macromolecules **32**, 2629 (1999).

[35] S. Rathgeber, T. Pakula, A. Wilk, K. Matyjaszewski, and K. L. Beers, J. Chem. Phys. **122**, 124904-1 (2005).

[36] W. Hu, J. Chem. Phys. **109**, 3686 (1998).

[37] A. K. Dasmahapatra, H. Nanavati, and G. Kumaraswamy, J. Chem. Phys. **127**, 234901 (2007).

**Table caption:**

**Table 1: Details of sample systems used in simulation with the branching parameters calculated from simulation (*g*) and by using equation 2 ($g_{th}$).**

**Figure captions:**

**Fig. 1. Schematic representation of H-shaped polymer. *N1* and *N2* represent the length of backbone and branches. Different color used for backbone and branch only for better visualization.**

**Fig. 2. Change in mean square radius of gyration, $\left\langle R_g^2 \right\rangle$, as a function of *B* for homopolymer and H-polymers with *N* = 512. The lines joining the points are meant only as a guide to the eye.**

**Fig. 3. Change in mean square radius of gyration (scaled with *N*) as a function of *B* for (a) homopolymer and (b) H-polymers with a series of *N*. The theta point is determined as the point where $\left\langle R_g^2 \right\rangle / N$ is equal for all *N*. The theta value of *B* is as indicated. The lines joining the points are meant only as a guide to the eye.**

**Fig. 4. Comparison of the density as a function of *B* for H-polymers with a series of *N*. The lines joining the points are meant only as a guide to the eye.**

**Fig. 5. Change in (a) mean square branch to end distance, (b) mean square radius of gyration of backbone and (c) scaled radius of gyration of the entire H-polymers as a function of *B* for a series of H-polymers. The lines joining the points are meant only as a guide to the eye.**

**Fig. 6. Change in (a) mean square branch to end distance and (b) mean square radius of gyration of backbone as a function of branch length (*N2*) for a series of H-polymers with *N1* = 32, 64, 128 and 256 at *B* = 0.036. The lines joining the points are meant only as a guide to the eye.**

**Fig. 7. Change in (a) mean square branch to end distance and (b) mean square radius of gyration of backbone as a function of branch length (*N2*) for a series of H-polymers with *N1* = 32, 64, 128 and 256 at *B* = 0.1. The lines joining the points are meant only as a guide to the eye.**

**Fig. 8. Change in mean square radius of gyration as a function of *B* for H-polymers with constant *N*: (a) 160 and (b) 192, by varying *N1* and *N2*. The lines joining the points are meant only as a guide to the eye.**

**Fig. 9. Change in mean square radius of gyration as a function of *B* for H-polymers with equal length of backbone and branch (*N1* = *N2*): (a) 32 and (b) 64. The lines joining the points are meant only as a guide to the eye.**

**Fig. 10. Change in mean square radius of gyration of branch $\langle R_g^2 \rangle_{N2}$ as a function of *B* for a series of H-polymers with (a) *N2* = 32 and (b) *N2* = 64. The lines joining the points are meant only as a guide to the eye.**

**Fig. 11. Snapshots of partially collapsed (viz., crumpled globule) structures at *B* = 0.036 for a series of H-polymers: (a) 64_16_128, (b) 64_32_192, (c) 64_48_256, (d) 64_64_320**

**and (e) 64_80_384, showing that the backbone is more collapsed than branches, which are more solvated and possess higher chain dimension than backbone. Black and magenta color used for backbone and branch units for better visualization.**

**Fig. 12. Snapshots of collapsed structures at *B* = 0.1 for a series of H-polymers: (a) 64_4_80, (b) 64_16_128, (c) 128_32_256 and (d) 128_96_512. These structures resemble to "sandwich", wherein backbone units are flanked by side branches from either end. Black and magenta color used for backbone and branch units for better visualization.**

**Fig. 13. Snapshots of collapsed structures at *B* = 0.1 for a series of H-polymers: (a) 64_32_192, (b) 64_64_320, (c) 64_80_384 and (d) 64_96_448. These structures resemble to "Janus" morphology. Black and magenta color used for backbone and branch units for better visualization.**

**Fig. 14. Change in asphericity factor, $\delta^*$ as a function of *B* for (a) homopolymer and (b) H-polymers with *N* = 64, 128, 256 and 512. The lines joining the points are meant only as a guide to the eye.**

**Fig. 15. Change in asphericity factor, $\delta^*$ as a function of *B* for H-polymers (a) for a series of branch length with a backbone length 64; inset: $\delta^*$ vs. *N2* at *B* = 0.1, (b) for a series of backbone length with a branch length 64, and (c) linear and H-polymers with *N* = 512. The lines joining the points are meant only as a guide to the eye.**

**Table 1**

| Sl. no. | *N1* | *N2* | *N* | $\lambda$ | $g$ | $g_{th}$ |
|---|---|---|---|---|---|---|
| 1 | 16 | 4 | 32 | 0.5 | 0.86 | 0.86 |
| 2 | 32 | 8 | 64 | 0.5 | 0.88 | ,, |
| 3 | 64 | 16 | 128 | 0.5 | 0.83 | ,, |
| 4 | 128 | 32 | 256 | 0.5 | 0.89 | ,, |
| 5 | 256 | 64 | 512 | 0.5 | 0.88 | ,, |
| 6 | 96 | 24 | 192 | 0.5 | 0.88 | ,, |
| 7 | 64 | 48 | 256 | 0.25 | 0.73 | 0.74 |
| 8 | 128 | 96 | 512 | 0.25 | 0.75 | ,, |
| 9 | 64 | 80 | 384 | 0.167 | 0.71 | 0.70 |
| 10 | 64 | 64 | 320 | 0.2 | 0.70 | 0.71 |

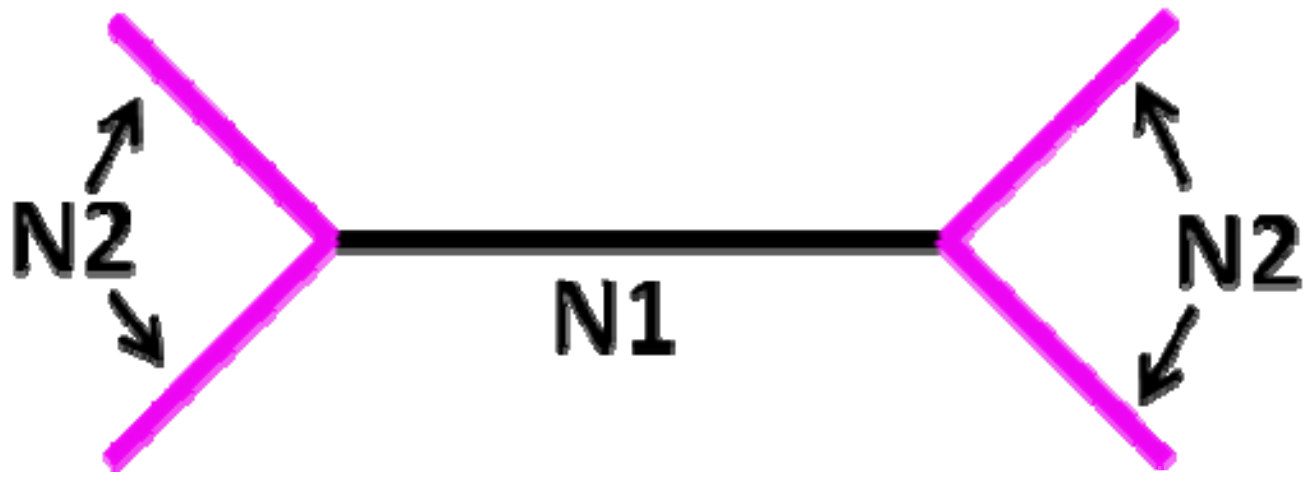

**Figure – 1**

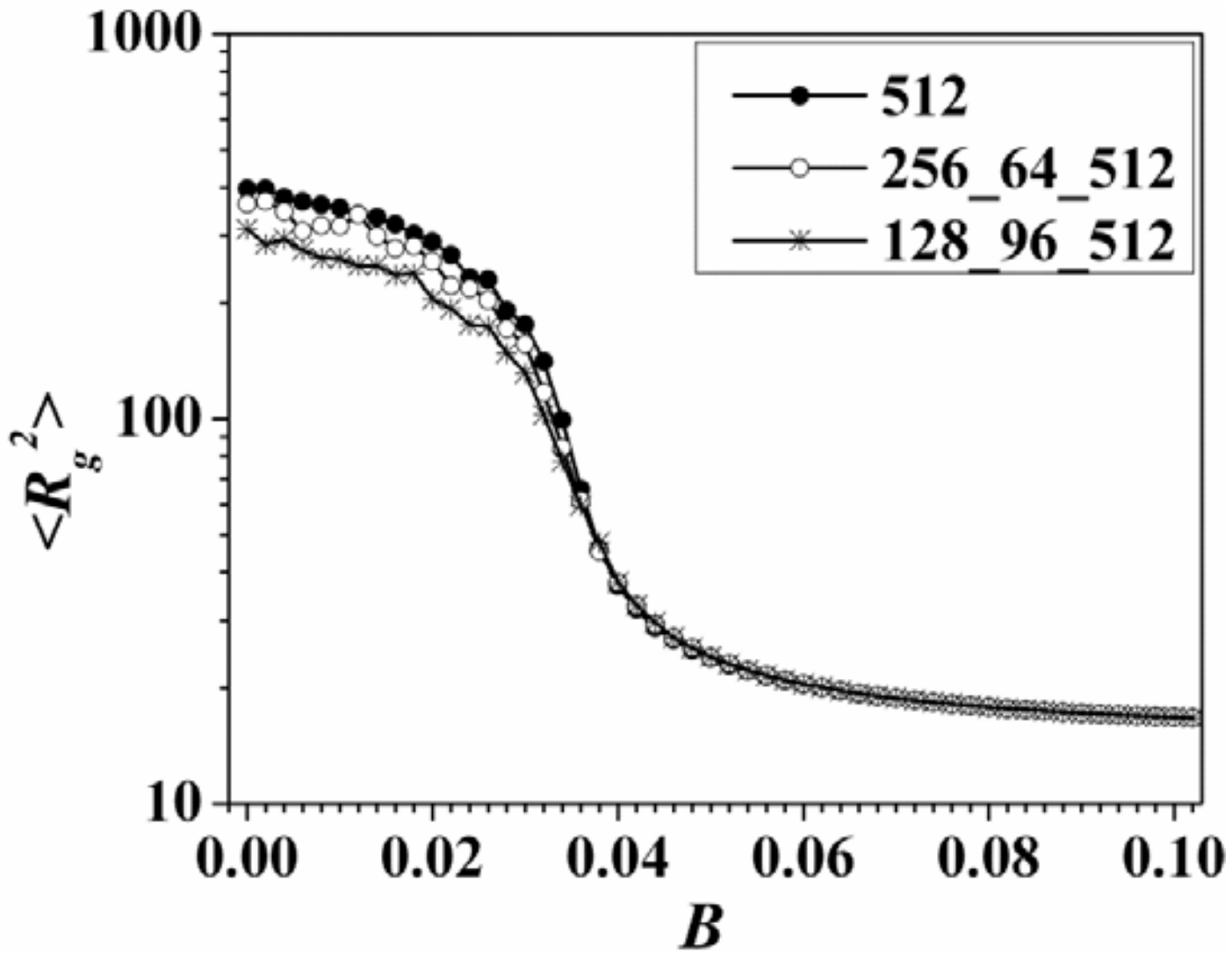

**Figure – 2**

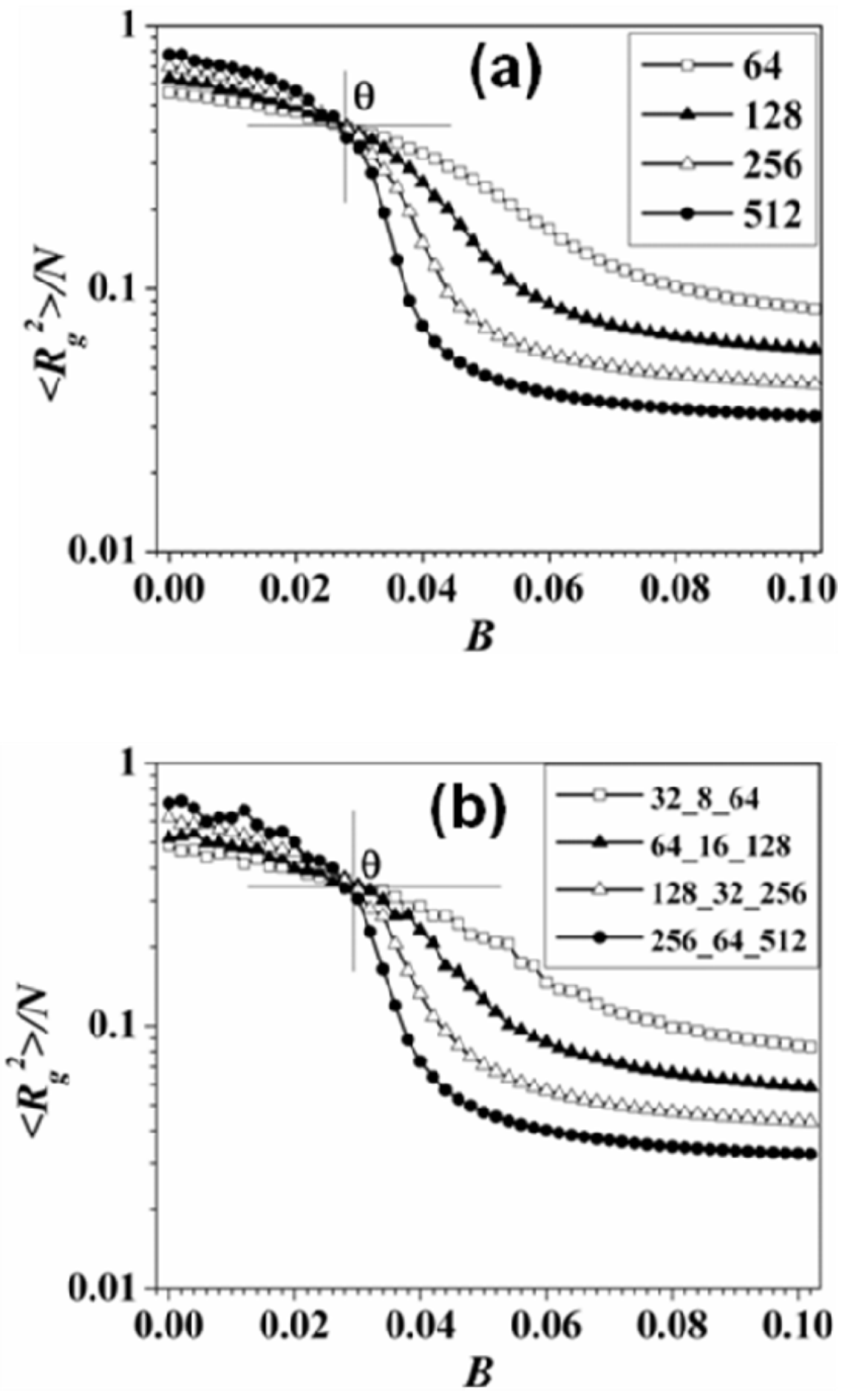

**Figure – 3**

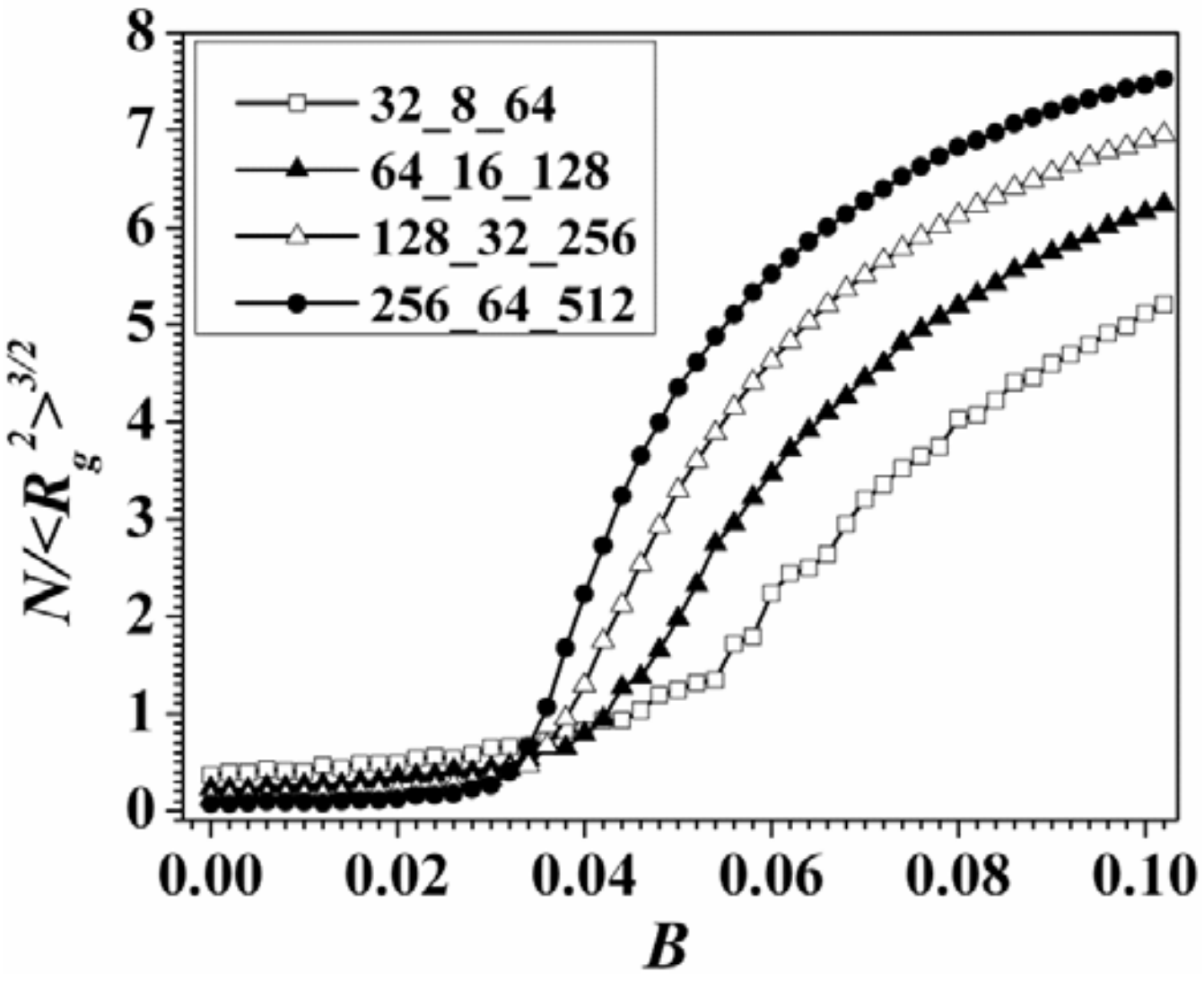

**Figure – 4**

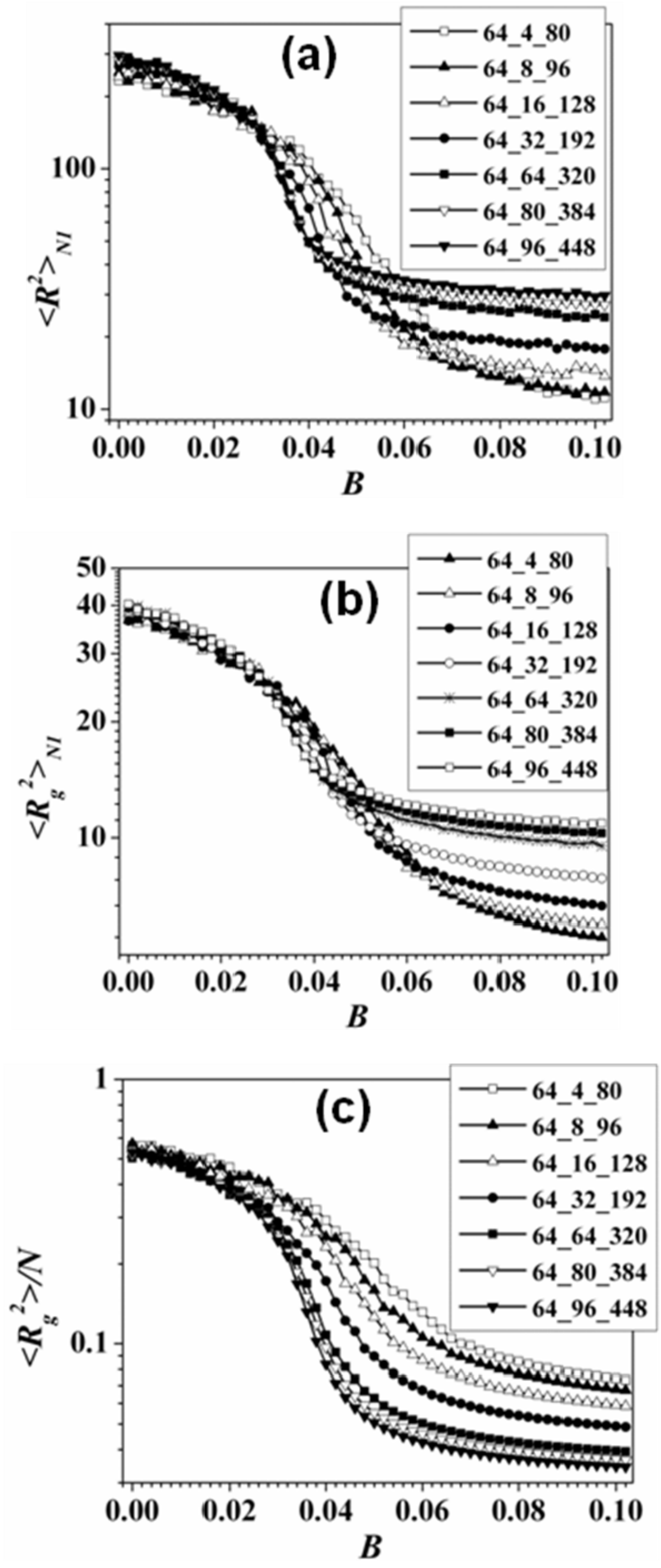

**Figure – 5**

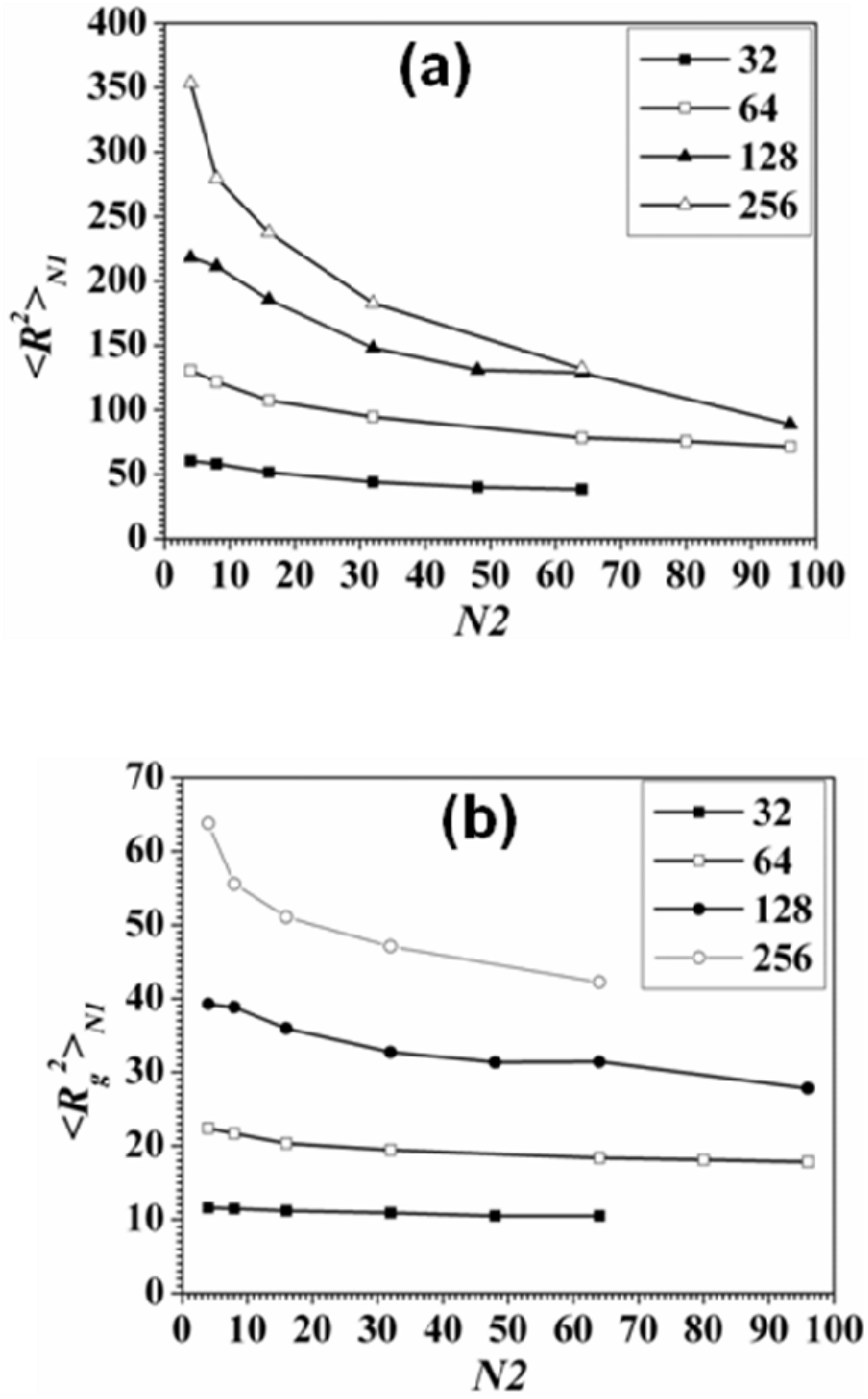

**Figure – 6**

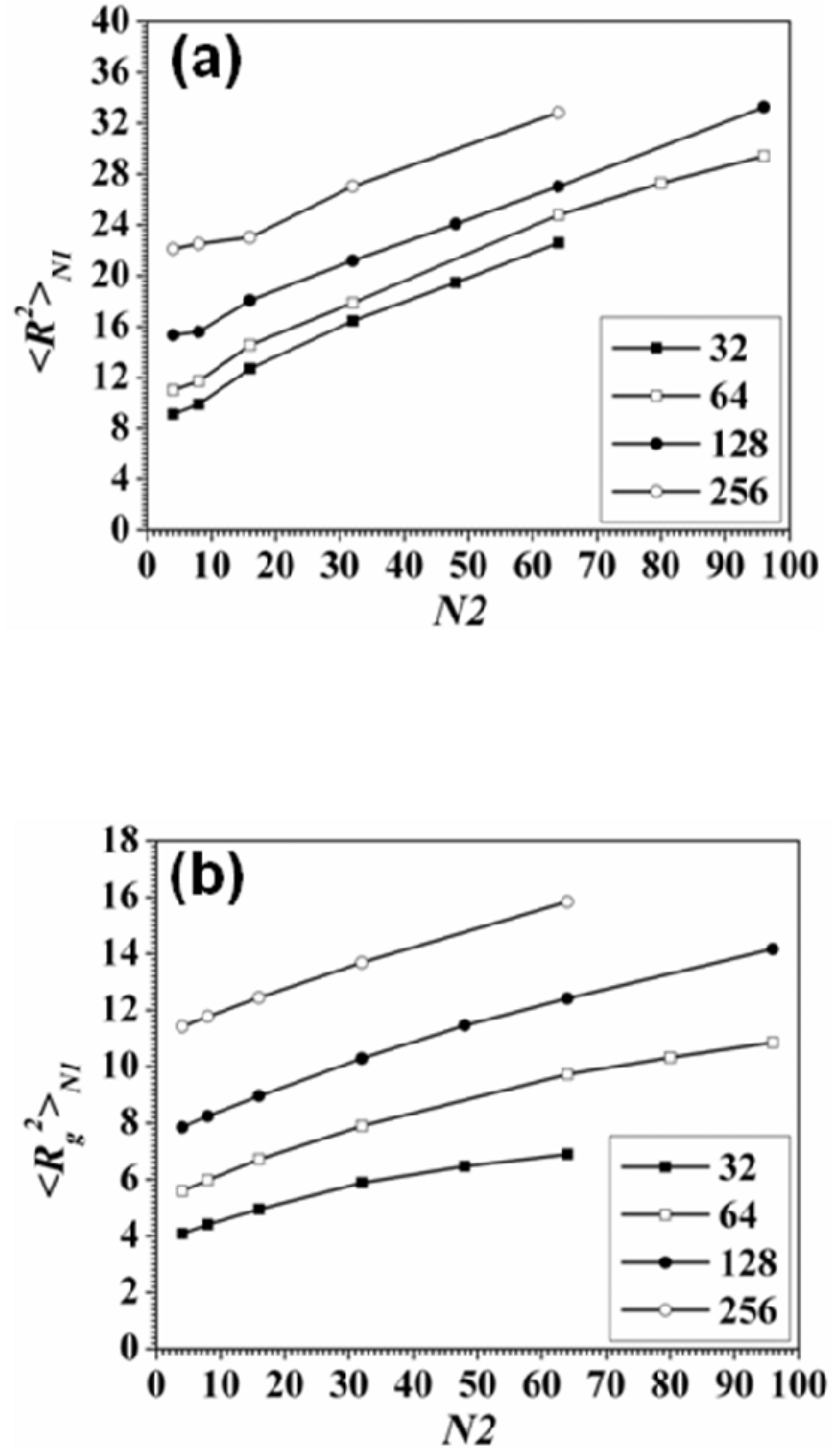

**Figure – 7**

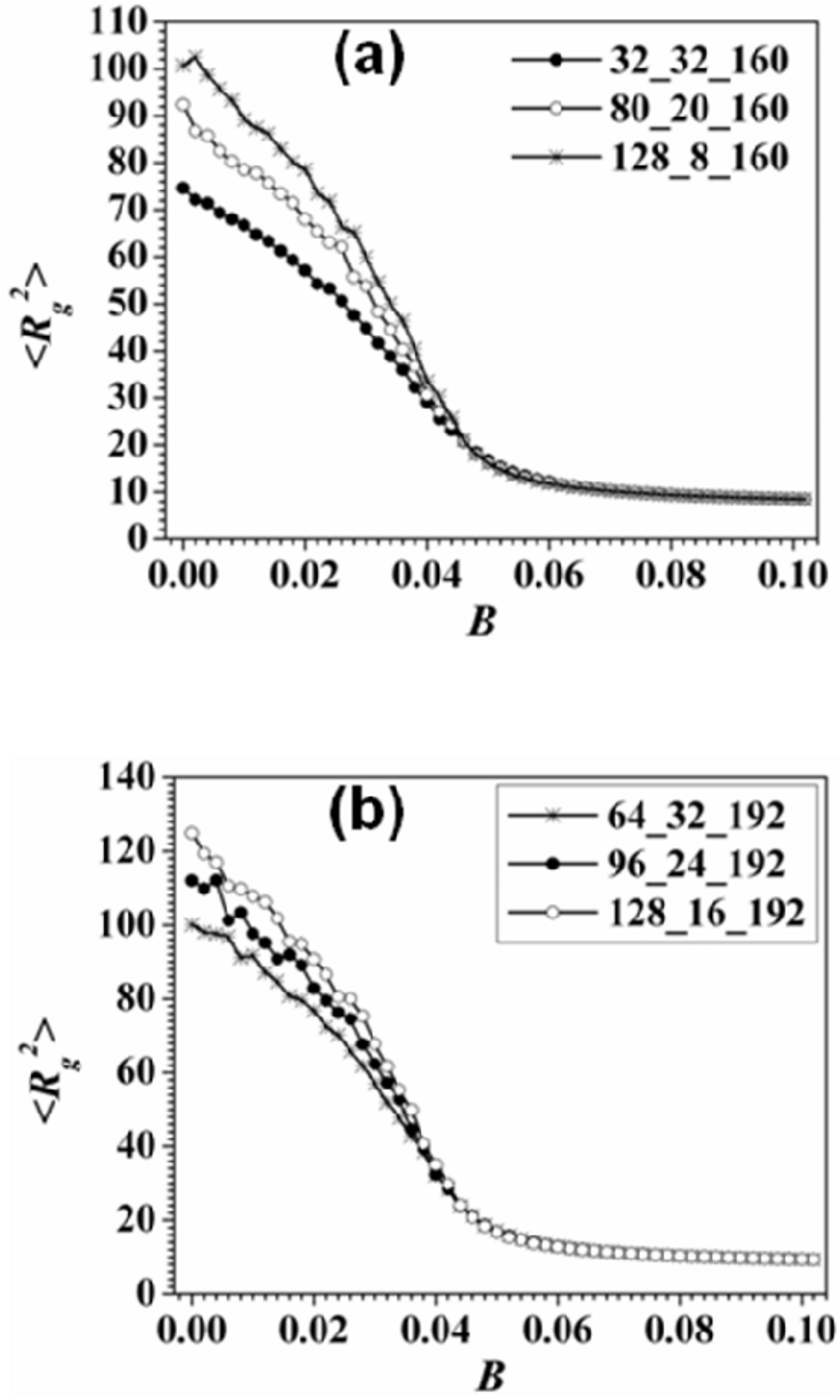

**Figure – 8**

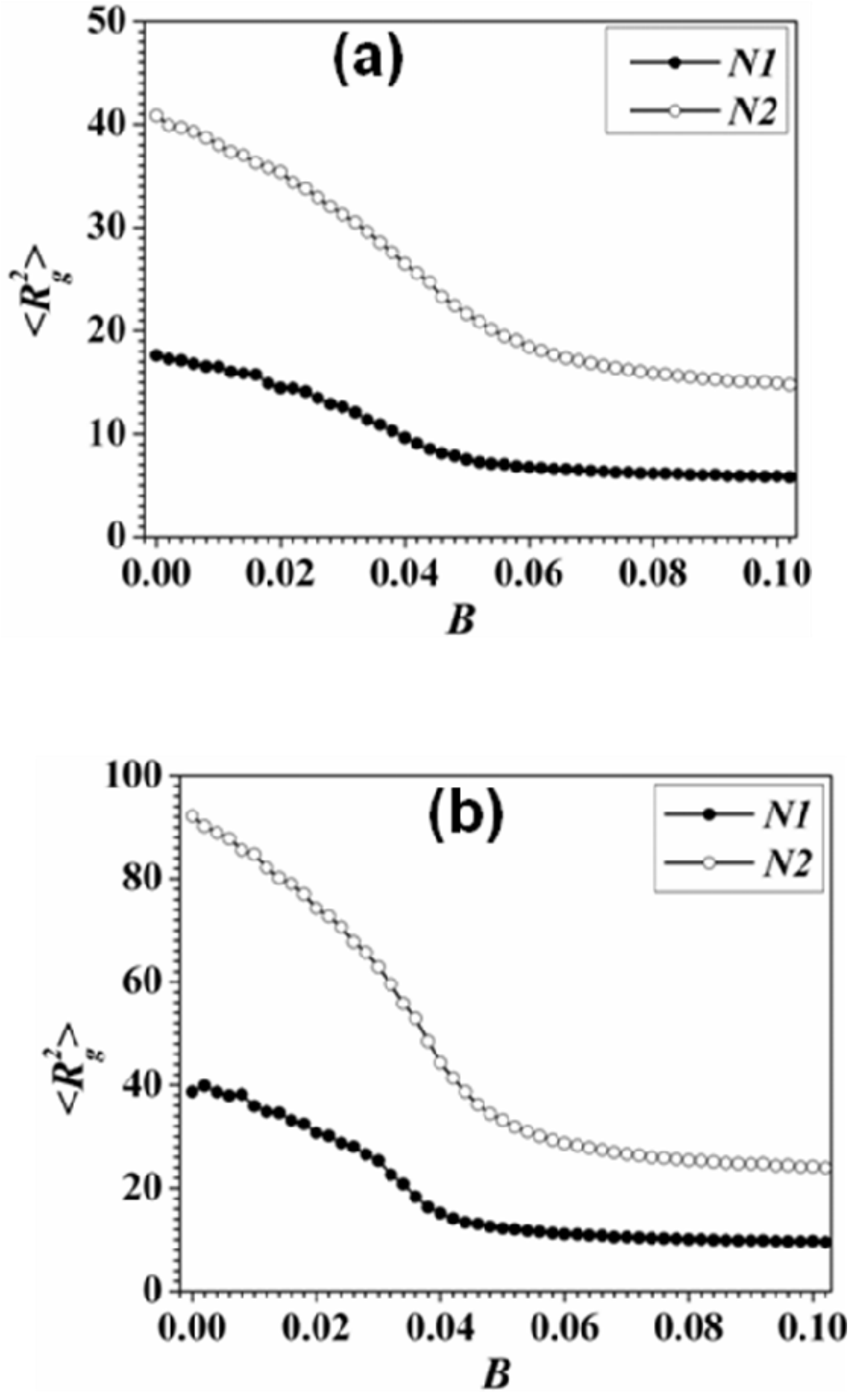
(a)
N1
N2
<$R_g^2$>
B
0.00 0.02 0.04 0.06 0.08 0.10
0 10 20 30 40 50
(b)
N1
N2
<$R_g^2$>
B
0.00 0.02 0.04 0.06 0.08 0.10
0 20 40 60 80 100

**Figure – 9**

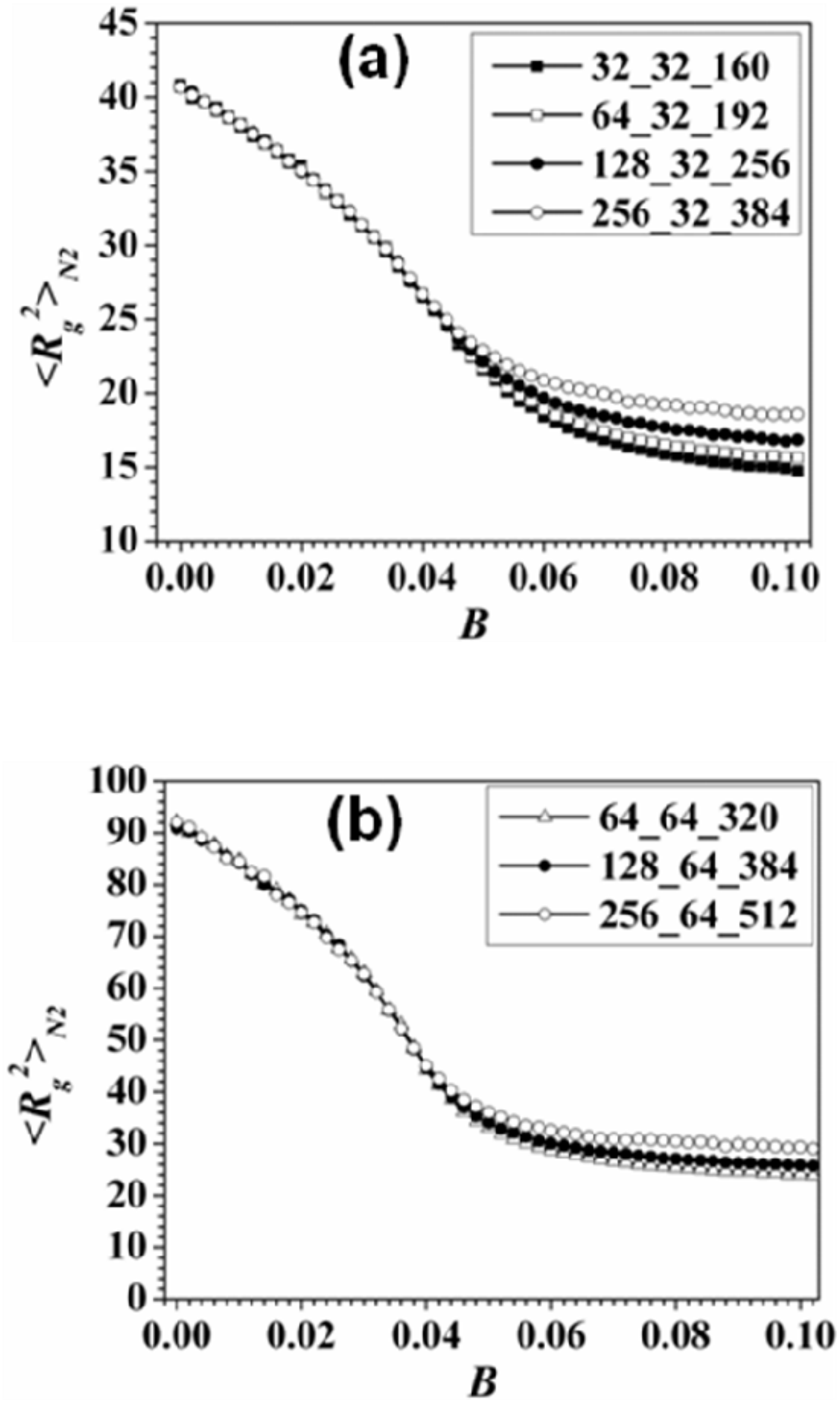

**Figure – 10**

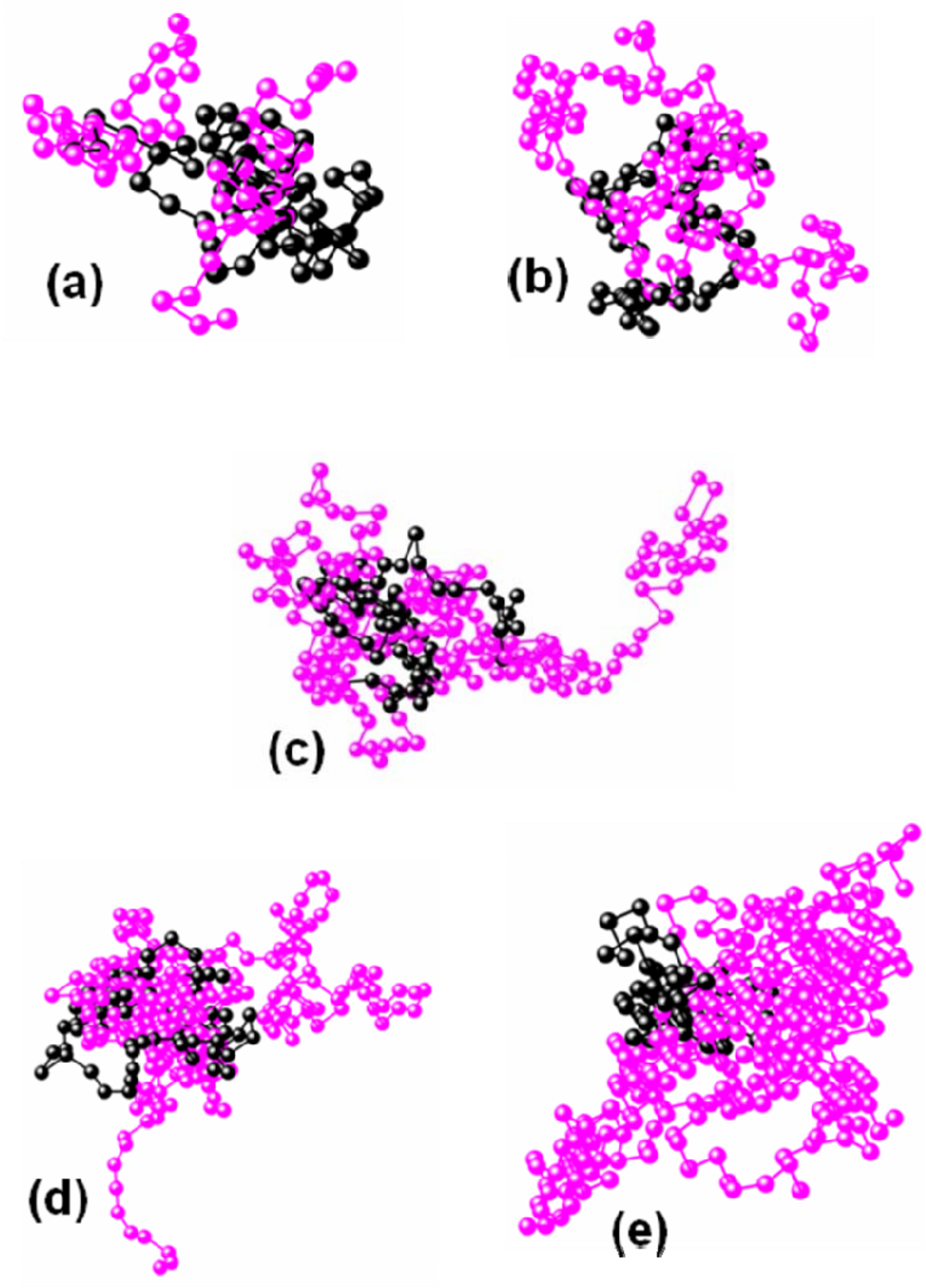

**Figure – 11**

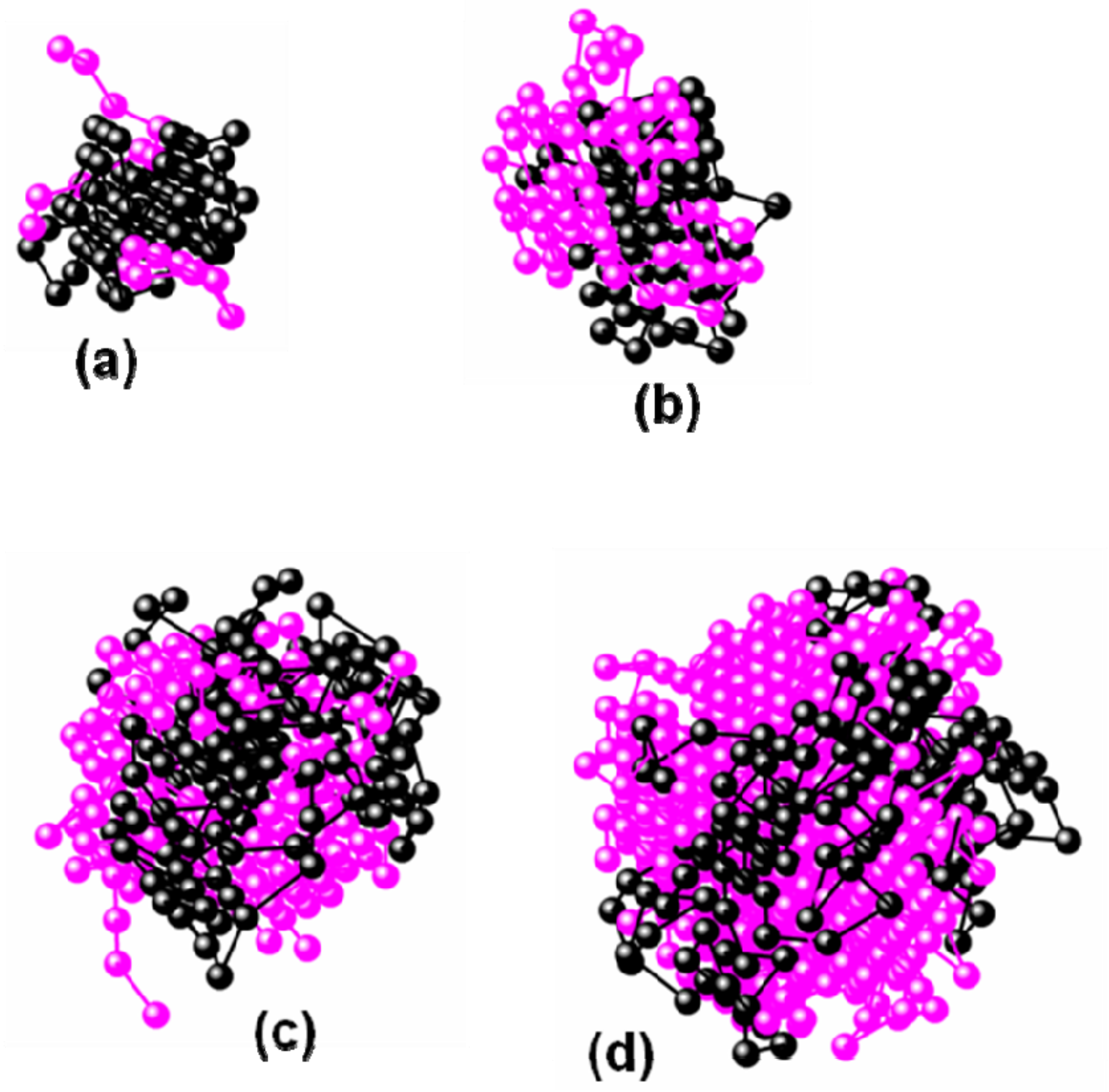

**Figure – 12**

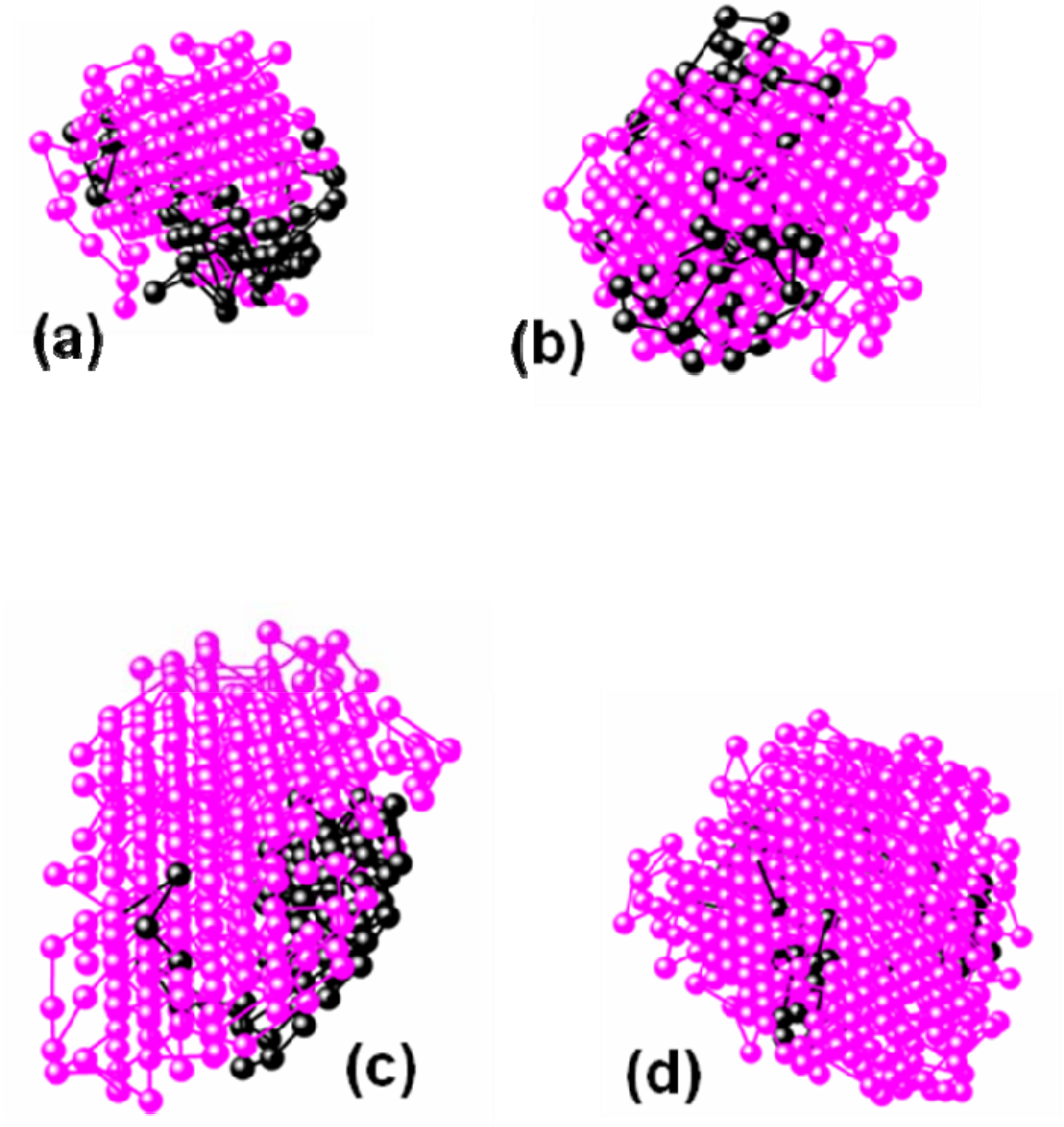

**Figure – 13**

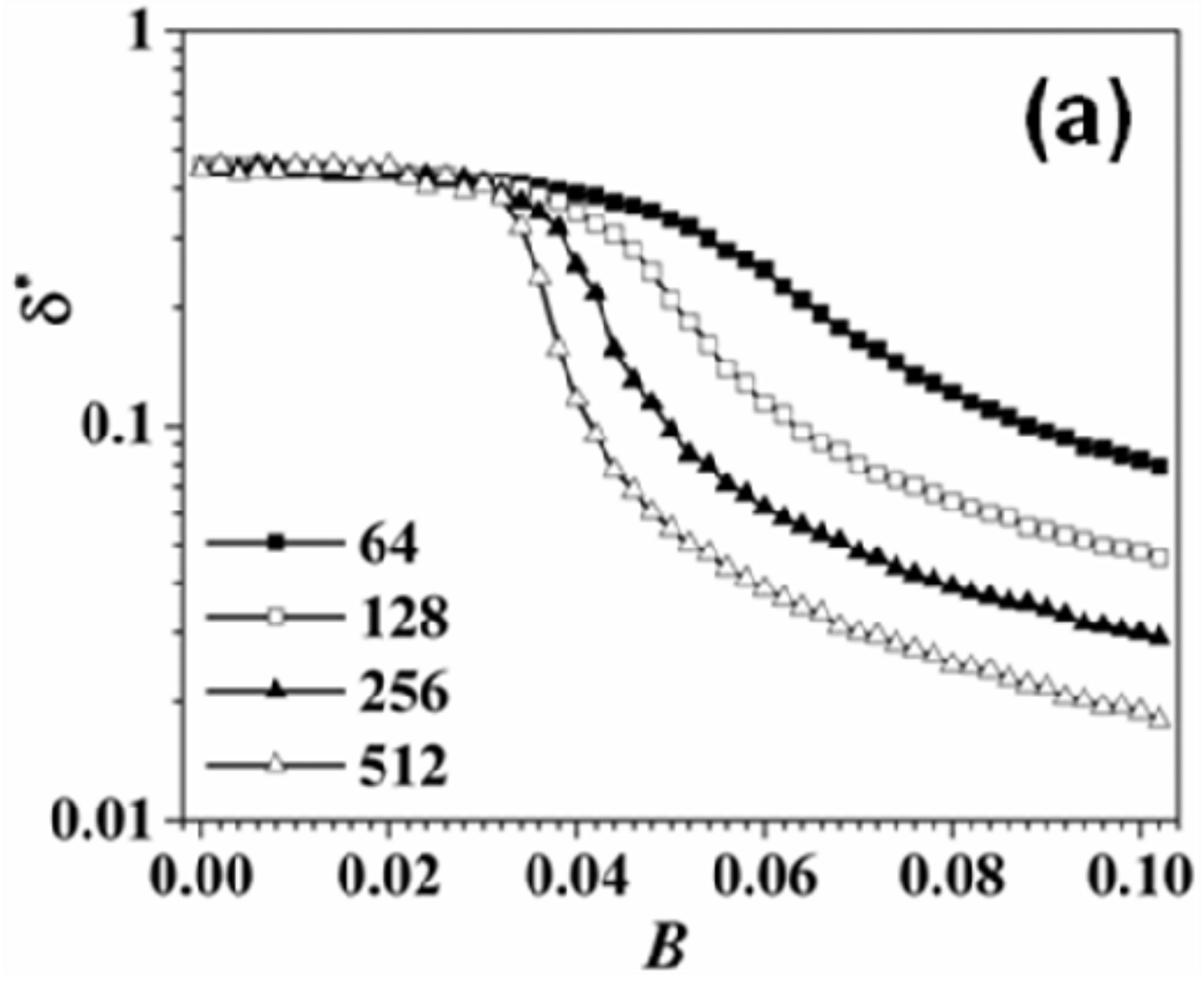

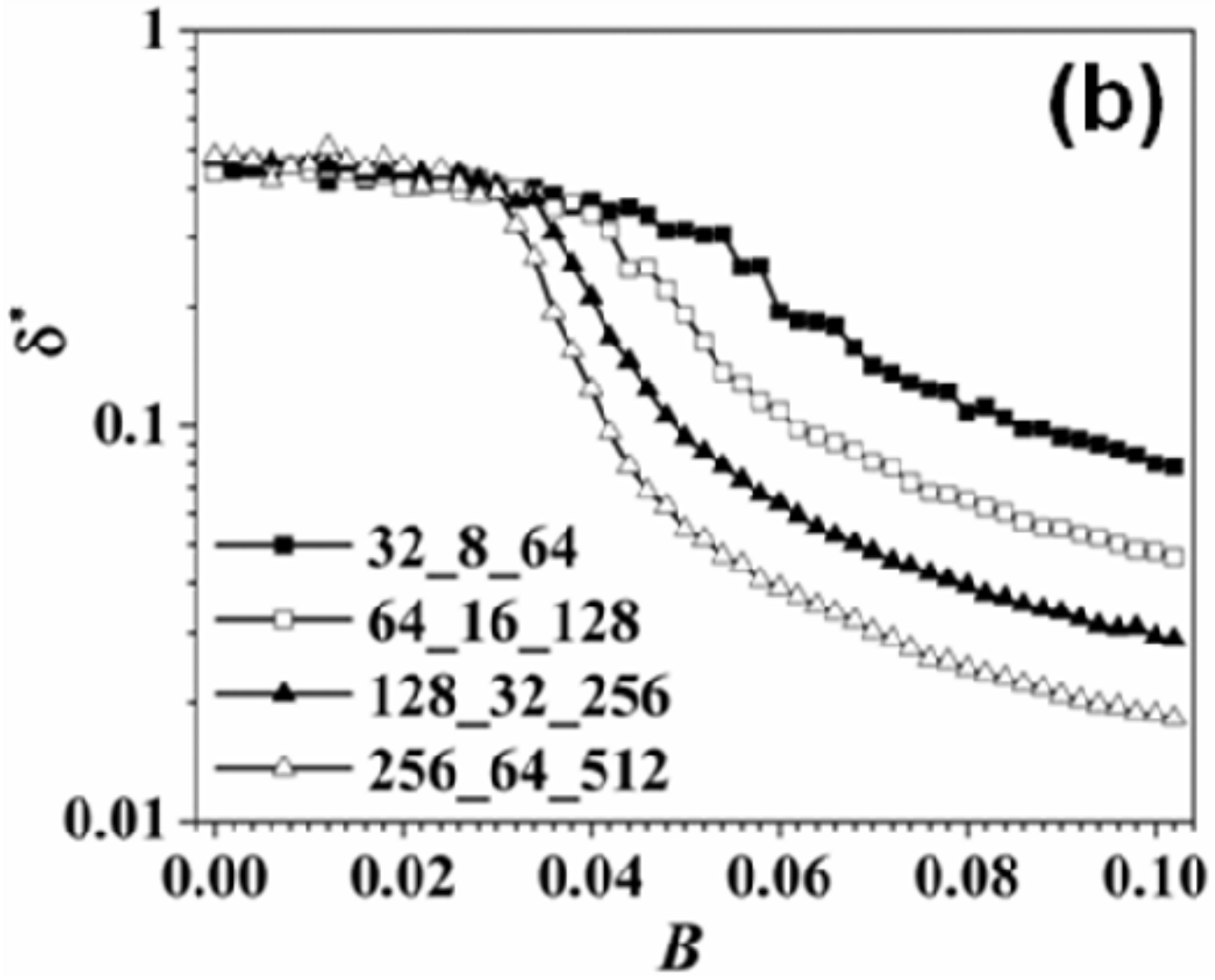

**Figure – 14**

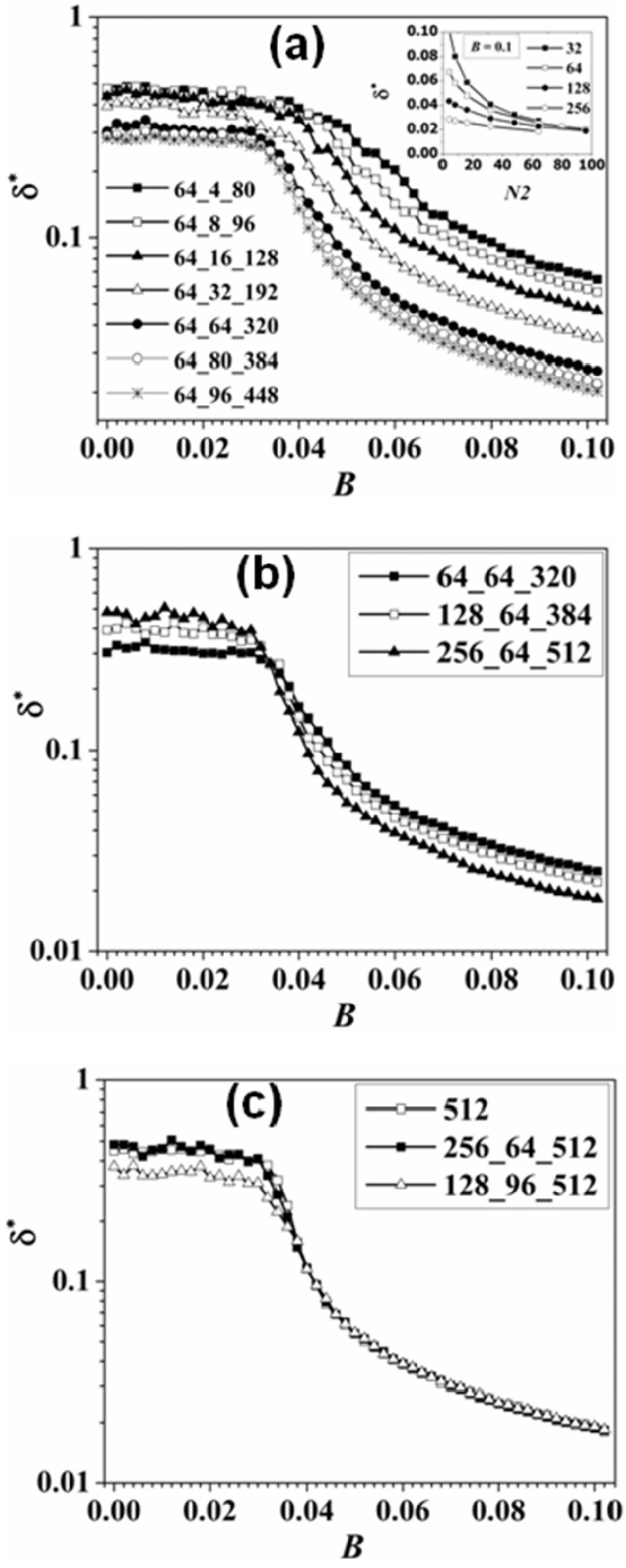

**Figure – 15**